\documentclass[prl,showpacs,twocolumn,showkeys,amssymb,amsmath,aps]{revtex4}
\usepackage{graphicx}% Include figure files
\usepackage{bm}% bold math
\begin{document}
\title{An on-the-fly approach optimized switching: A special case of  \\
 free energy simulation under nonequilibrium feedback control}
\author{M. Ponmurugan\footnote{mail address : mpn@imsc.res.in}} 
\affiliation{The Institute of Mathematical Sciences, 
C.I.T. Campus, Taramani, Chennai 600113, India}

%\date{\today}

\begin{abstract}

We discuss the optimized switching free energy simulations
in an analogy with the systems which are driven under 
nonequilibrium feedback control.
We find an on-the-fly simulation approach of  
switching optimization is a special case of  
the nonequilibrium process under feedback control.   
In this approach, 
the switching rate is allowed to vary during 
the simulation and the optimization is done on-the-fly 
by utilizing the part of the simulation outcomes as the feedback 
information. In such a case, 
one should use generalized Jarzynski equality under nonequlilibrium 
feedback control for  
free energy calculation instead original Jarzynski equality.

\end{abstract}
\maketitle

There are very few relations in statistical dynamics 
has been used to calculate the equilibrium thermodynamics 
properties for the systems which are driven arbitrarily 
far-from-equlibrium \cite{jar,crooks,other}.
One of these is the Jarzynski equality \cite{jar} which relates 
nonequilibrium measurements of the work done on the system 
to equilibrium free energy differences.
The second law of thermodynamics can be quantitatively
described by the fluctuation theorem  which are closely 
related to the Jarzynski equality \cite{jar,crooks}.
A system initially at equilibrium with temperature (inverse) $\beta=1/k_BT$
is externally driven from its initial state to 
final state by nonequilibrum 
process satisfies the detailed fluctuation theorem, 
\begin{eqnarray}
\frac{P(W)}{P(-W)} &=& e^{\beta (W- \Delta F)}
\end{eqnarray}
and its integrated version, the Jarzynski equality
\begin{eqnarray}
 \langle e^{-\beta W} \rangle &=& e^{-\beta \Delta F}.
\end{eqnarray}
Where $W$ denotes the work performed on the system, $\Delta F$ 
is the free energy  difference of the system between its final 
and initial equilibrium states and $P(\pm W)$ is the work probability 
distribution in forward (+) and reverse (-) direction. 
This relationship is widely used in experiments \cite{lip,col} as well as 
simulations \cite{steer} in many branches of Science (see, eg.\cite{van}).

Various experiments and simulations has been performed by
adopting a suitable time-dependent driving scheme 
described by an external control switching  protocol.
Even though the Jarzynski equality is valid for    
any time-dependent driving scheme, 
the efficiency of a nonequilibrium switching simulation 
which use Jarzynski equality to estimate precise free energy difference
is depends on the switching function. A well selected switching function can 
significantly minimize the associated dissipated work $W_d=W-\Delta F$ 
and reduce the computational cost of nonequilibrium 
free energy simulation \cite{seif,seif1,engel}.

In simulations, one has to search for the switching protocol 
that minimizes the mean work required to drive the system 
from one given equilibrium state to another in 
finite time  such that the exponential average 
should provides precise free energy estimates \cite{seif,seif1,engel}.
Since the optimized switching protocol minimizes the average work \cite{engel}, 
most nonequilibrium free energy simulations which use 
Jarzynski equality employ optimized switching function \cite{kon}.
One simple and reasonable method of optimization   
utilizes  the informations obtained from an ensemble 
of short simulation which are carried out at
relatively low switching speeds, usually within 
the linear response regime \cite{kon}. 
Once the optimized control switching parameter has been estimated,
the long run free energy simulations then performed with 
the optimized switching function at a slower speed.

In a recent paper \cite{onfly},
an on-the-fly approach is proposed for estimating 
an efficient switching function during a single nonequilibrium 
free energy simulation. In this approach,
the part of the simulation carried out by 
allowing the switching function 
to vary and the remaining part continued  
with the optimized switching function which  has been estimated  
on-the-fly. In this approach, the informations 
obtained from the earlier part of the
simulation outcomes are used to estimate the 
optimized switching function.
Here, the informations used for optimization are the 
switching rate, the variance of 
the simulation outcomes and its autocorrelation function \cite{kon,onfly}.
It has been shown very recently that the accuracy of the free 
energy estimates also depends on shape of 
the switching protocol \cite{dell}.

In most of the optimization problems \cite{kon,onfly,dell},  
the (feedback) informations about the physical system plays 
crucial role in a direct or indirect way to
calculate the optimized switching  functions.
This can provide the natural link between the 
switching optimization and the informations.    
In this paper, we discuss the optimized switching free 
energy simulations in an analogy with the systems which are 
driven under nonequilibrium feedback control \cite{feedref} and
argue that the work value obtained from 
an on-the-fly approach in general does not  
satisfy Jarzynski equality.

The evolution of the physical systems can be modified or controlled 
by repeated operation of an external agent called controller \cite{feedref,frev}.
In contrast to open loop controller which operates on the 
system blindly, the feedback or closed loop controllers use 
information about the state of the system. The feedback
is the process performed by the controller of measuring the 
system, deciding on the action given the measurement output, 
and acting on the system \cite{cao}. For example, in a single molecule
Atomic Force Microscopy experiment, the external agent 
is an electric feedback circuit which detects the motion of  
the cantilever and manipulate 
the control force proportional to its velocity \cite{qian}. 
The proper utilization of the information about the 
state of the system in feedback control effectively 
improves the system performance \cite{feedref,frev,cao,qian,sagab1}. 
However, the presence of feedback control in physical system modifies both the 
Jarzynski equality and the fluctuation theorem \cite{qian}.

Recently, the Jarzynski equality is generalized 
to an experimental condition in which the system is driven between 
two equilibrium state via nonequilibrium process  
under forward feedback control \cite{sagaf}.
At a given time, the controller measure the 
partial state of the system. The result of the measurement 
determines the action the control will take. 
The additional information on the system 
provided by the measurement further 
determines the system states.
The equilibrium free energy difference for the driven  system (which 
locally satisfies detailed fluctuation theorem) under nonequilibrium 
feedback control can be calculated from the generalized Jarzynski 
equality \cite{sagaf}
\begin{eqnarray}\label {equl1}
\langle e^{- \sigma - I} \rangle &=& 1, 
\end{eqnarray}
where $\sigma = \beta (W - \Delta F)$ and $I$ 
is the mutual information measure obtained from the 
feedback controller \cite{sagaf}. The average is 
taken from the distribution of work in forward 
direction  with feedback control. The work distribution 
with feedback control is computed without the 
knowledge of internal details of  the controller \cite{sagaf}.

Let $P(W,I)=P(W|I)P(I)$ be the (joint) probability distribution
of work with feedback information. The above equality, Eq.(\ref{equl1}),
can be written as,
\begin{eqnarray}\label {equl2}
\int \int P(W|I) P(I) \ e^{- \sigma - I} \ dW \ dI  &=& 1,
\end{eqnarray}
where $P(W|I)$ is the conditional probability for obtaining
the outcome $W$ given the mutual information measure $I$.
Since the controller often measure the partial state of 
the system, the changes in 
feedback information measure \cite{cao,beck} occurs 
with probability $P(I)$.

The feedback control enhances our controllability of 
small thermodynamics systems \cite{feedref,frev,cao,qian,sagab1}. 
In free energy simulations this feedback mechanism 
can be very helpful for sampling rare trajectories
for precise free energy difference calculations. 
Since the work that  performed on a thermodynamic system 
can be lowered by feedback control \cite{sagab1,sagaf},
we can say that using proper feedback mechanism 
in a free energy simulation is equivalent to carry out 
an optimized switching free energy simulation
which minimizes the average work. 
In this perspective, we notice that an 
on-the-fly approach is a special case of nonequilibrium 
process under feedback control as follows.

In simulations, the free energy difference 
between the two equilibrium states can be calculated 
in general by pulling the sytem from one equilibrium state 
to another state along a switching path. 
The path connecting the two states will be parameterized  
using the variable $\lambda$, with $0 \le \lambda \le 1$.
The switching rate describes the nature of the switching 
process to be an equilibrium (infinitely slow) or nonequilibrium (fast).   
In an on-the-fly approach, up to the history length time, $t_h$ \cite{onfly},
the switching  function  parameter, $\lambda_t$, vary during the simulation.
The simulation outcomes obtained within $t_h$ are used 
as a feedback information to estimate the  
optimized switching function, $\lambda^{\star}_t$. The remaining part of the
simulation carried out with this fixed $\lambda^{\star}_t$. 
This is equivalent to say, 
at time $t_h$, the (un-known) controller \cite{note}
measure the partial state of the system and determines 
the optimized switching function $\lambda^{\star}_t$.
Once the switching function to be
optimized, the outcome of the remaining part (after $t_h$) of the   
simulation is performed  by fixed feedback control 
with feedback information measure, $I(\lambda^{\star}_t)$.

In this aspect, the feedback information in an on-the-fly approach  
can be characterized by the mutual information measure \cite{sagaf},
$I_1=0$ for $t \le t_h$ and  
$I_2=I(\lambda^{\star}_t)$ for $t>t_h$.
From a single measurement of the controller, at $t_h$
the mutual information measure $I(\lambda^{\star}_t)$ 
is obtained over all simulation outcomes for $t \le t_h$.  
The term mutual information is appropriate in the sense that
in this approach, the optimized switching function $\lambda^{\star}_t$
depends on the feedback information measure $I(\lambda^{\star}_t)$ which
in turn depends on $\lambda^{\star}_t$ for further simulation.
In an on-the-fly approach, the switching function is 
fixed at $\lambda^{\star}_t$ after $t_h$ which indicates that
the  controller measure the partial state of the 
system  only once (at $t_h$) for system control (switching optimization)
instead of  many times in general. In this sense, we 
can say that an on-the-fly approach is a special case of 
nonequilibrium process under feedback control.

It should be noted that, in systems with 
feedback control  the simulation/experimental outcomes 
are different from controller measurement outcomes.
A single outcome of the controller measurement is a result 
of analyzing many realizations of the 
simulation/experimental outcomes. That is the controller 
utilizes the part of the simulation/experimental outputs 
and calculate its measurement outcome, see ref.\cite{frev}.
In an on-the-fly approach, the controller utilizes all 
simulation outcomes for $t \le t_h$ and obtained the
optimized switching function $\lambda^{\star}_t$.  
Due to the single measurement of the controller at $t_h$,
work  outcomes of the simulation associated with   
mutual information measure $I_1=0$ for 
$t \le t_h$ and $I_2=I(\lambda^{\star}_t)$ for $t > t_h$. 
Since the equality, Eq.(\ref{equl1}), does not depends on the 
internal details of  the controller \cite{cao,sagaf}
and the controller made measurement only once at $t=t_h$,
Eq.(\ref{equl2}) can be written in this case as
\begin{eqnarray}
\gamma(I_1) \ P(I_1) + \gamma(I_2) \ e^{-I_2} \ P(I_2) &=& 1, 
\end{eqnarray}
where,
\begin{eqnarray}
\gamma(I_i) &=& \int P(W|I_i) \ e^{-\sigma} \ dW, \ (i=1,2).
\end{eqnarray}

In contrast to the  
original Jarzynski equality, $\langle e^{- \sigma} \rangle =1$, 
if we measure $\langle e^{- \sigma} \rangle$ in the 
presence of the feedback control, the calculated value 
is differ from unity \cite{sagaf} which is given as 
\begin{eqnarray}
 \langle e^{-\sigma} \rangle &=& \gamma, 
\end{eqnarray} 
where $\gamma$ is a measure which characterizes the 
efficacy of the feedback control \cite{sagaf}.
Hence, in an on-the-fly approach, the free energy difference
obtained from  original Jarzynski 
equality \cite{onfly} is in general inappropriate.
In this approach one should use the generalized 
Jarzynski equality with proper feedback information measure 
$I(\lambda^{\star}_t)$ provided the simulation process 
should satisfy the local detailed balance or 
the detailed fluctuation theorem \cite{sagaf}.

Since the switching function under feedback control  
depends on the controller measurement outcomes \cite{sagaf},
if we perform nonequilibrium  simulation under 
feedback control, the different trajectories will have 
different switching function. These kind of
different switching function has been incorporated
in an  on-the-fly approach simulation \cite{onfly} 
clearly confirm that  this approach is a special case of 
free energy simulation with feedback control.
Thus, the work value obtained from an on-the-fly
approach in general does not satisfy 
original Jarzynski equality. The un-noticed sign of 
violation of original Jarzynski equality for those 
simple systems studied in an on-the-fly approach \cite{onfly} 
is in question. In fact,  
the free-energy estimate by the original Jarzynski equality 
for those systems studied in an on-the-fly approach
seems to be pretty good which can makes a general impression 
that the generalized Jarzynski equality does not needed. 
The main motivation for our analogy argument of on-the-fly approach 
with nonequilibrium process under feedback control  
is that the stated conclusion of the un-noticed sign of violation 
of original Jarzynski identity in an on-the-fly approach 
is not definite for all systems (see, ref. \cite{qian}), particularly, 
the folding and unfolding free energy simulations studies for
complex macromolecular systems.
It would be interesting to carry such an
investigation on macromolecular systems which has many 
folding intermediates \cite{vanes}.   
Finally, If one can carry on-the-fly approach simulation studies on complex systems
and insist to use original jarzynski equality for free enregy estimates,
within the confident level of certain hypothesis \cite{pande}   
one should properly test the violation of original Jarzynski equality.

In conclusion, our analogy argument of the switching 
optimization with nonequilibrium process under feedback control
indicates that an on-the-fly  approach switching simulation  
is a special case of a nonequilibrium  
free energy simulation with  feedback control. 
In such a case, using  
generalized Jarzynki equality under nonequilibrium 
feedback control is 
appropriate for free energy difference calculation
instead original Jarzynki equality.


\begin{thebibliography}{100}
\bibitem{note}
In an on-the-fly approach, the arbitrary controller is one who
carry optimized switching calculation.
\end{thebibliography}
\end{document}